\documentclass[aps,pre,superscriptaddress,floats,psfig]{revtex4}
\usepackage{graphicx}
\usepackage{epsfig}
\usepackage{psfig}
\usepackage{dcolumn}
\usepackage{bm}
\usepackage[latin1]{inputenc}
\usepackage{amsmath}
\usepackage[dvips]{color}

\begin{document}

\title{Propagation of finite amplitude electrostatic disturbances \\
in a magnetized Vlasov plasma}
\author{Maurizio~Lontano}
\address{Istituto di Fisica del Plasma, C.N.R.,
EURATOM-ENEA-CNR Association, Milan, Italy}
\author{Laura~Galeotti}
\address{Dipartimento di Fisica, Universit\`a di Pisa,
Pisa, Italy}
\author{Francesco~Califano$^{1,2}$}

\begin{abstract}
\noindent
A 1D2V open boundary Vlasov-Ampere code has been implemented with 
the aim of making a detailed investigation of the propagation
of finite amplitude electromagnetic disturbances in an inhomogeneous
magnetized plasma. The code is being applied to study the propagation of 
an externally driven electromagnetic signal, localized
at one boundary of the integration interval, through a given
equilibrium plasma configuration with inhomogeneous plasma density and
magnetic field.
\end{abstract}

\maketitle

\section{Introduction}

Self-consistent electromagnetic electromagnetic fields in 
spatially non-uniform plasmas represent
one of the fundamental aspects of plasma physics with several
implications both in microwave and laser based experiments. Since the
`70es extensive theoretical
\cite{morales74,albritton75,deneef77a,colunga85}
and experimental
\cite{kim74,wong75,deneef77b}
investigations have been devoted to ponderomotive effects, particle
acceleration, wavebreaking, resonant absorption in plasmas with density
gradients. The problem is relevant for magnetized plasmas, as well,
when Bernstein waves are excited as a consequence of mode conversion
close to the hybrid plasma resonances. Recently, a renewed interest
for electron Bernstein wave physics has apperaed due to the
possibility of implementing an attractive radiation emission diagnostic in
fusion plasmas
\cite{efthimion99,chattopadhyay02,preinhaelter03}.
Then it is interesting to investigate the kinetic aspects of the
propagation of electromagnetic as well as electrostatic fields in a non
uniform plasma with arbitrary density and magnetic field scales, and
electric field amplitudes.

A 1D2V open boundary Vlasov-Ampere code has been implemented with the aim
of achieving a more realistic investigation of the propagation
of a finite amplitude signal in an inhomogeneous plasma, and it 
has been applied to study the propagation of an externally driven, 
localized charge density fluctuation in an unmagnetized plasma with 
an equilibrium spatially nonuniform density \cite{lontano04}.
Previous analyses of magnetized plasmas \cite{califano03,marchetto03a},
carried out in a slab geometry with periodic boundary conditions,
have shown that strongly anisotropic distribution functions 
are produced both in the electron \cite{califano03} and in 
the ion \cite{marchetto03a} populations during the interaction 
of plasma with an externally applied propagating electrostatic wave. 
In those cases the background plasma inhomogeneities could
be modelled by varying the ratio of the pump frequency to 
the electron plasma frequency \cite{marchetto03b}.

Here, the propagation of the spatially localized finite 
amplitude electromagnetic perturbation in a homogeneous 
magnetized collisionless plasma is preliminarly
investigated, the final aim being the study of 
electromagnetic fields propagating in a given equilibrium 
plasma configuration where both plasma density and magnetic 
field are inhomogeneous. The solution of the Vlasov equation, 
coupled with the Ampere equation, provides the electron 
distribution function and the electrostatic field
in the whole spatial range $0 < x < L$, at any time.
Results are presented relevant to moderate amplitude 
electromagnetic field perturbations.

\section{The physical model}

\begin{figure}[!h]
\hspace*{-1cm}\centerline{\psfig{figure=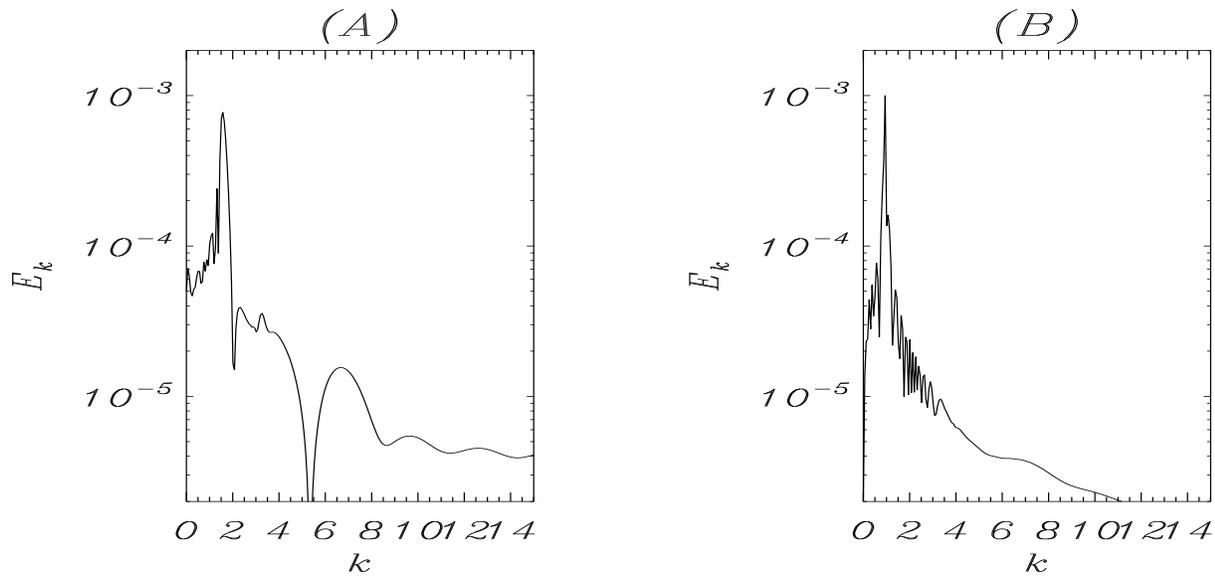,height=8cm,width=18.cm}}
\caption[ ]{The {\it k}-spectra of the electric field for $\omega = 2.1$, 
frame (A), and $\omega = 2.1$, frame (B).} 
\label{fig1}
\end{figure}

Let us consider a one dimensional magnetized plasma, localized in the
region $0 > x > L$, homogeneous in the plane $(y,z)$, with unperturbed
density $n_0 = n_{e0} = Z n_{i0}$, and magnetic field ${\bf B}_0 = B_0
\hat {\bf e}_z.$ The relevant dimensionless non relativistic Vlasov equations take the
form
\begin{equation}
{{\partial f_a}\over{\partial t}} + v_x {{\partial f_a}\over{\partial
x}} - \mu_a \left [ E_x^{tot} + v_y B_z \right ]
{{\partial f_a}\over{\partial v_x}} -
\mu_a \left [ E_y^{tot} - v_x B_z^{tot} \right ]
{{\partial f_a}\over{\partial v_y}}
= 0 ,
\label{1}
\end{equation}
where $a = e,i$, $\mu_e = 1$, and $\mu_i = -Z m_e / m_i$. Normalized
variables are defined as follows: $t \rightarrow t \omega_{pe}$, $x
\rightarrow x/d_e$, $v \rightarrow v/c$, $E (B) \rightarrow e E (B)
/m_e c \omega_{pe}$. Moreover, $\omega_{pe} = \left ( 4 \pi n_0 e^2/
m_e \right ) ^{1/2}$, $d_e = c/\omega_{pe}.$ As a result, in our units, 
the normalized length scale is the electron skin depth $d_e = 1$, while 
the (dimensionless) electrostatic length scale, the Debye length, is 
equal to the normalized thermal velocity, $\lambda_D = v_{th,e} / c$.  
In Eq.(\ref{1}) the components of the electric field are
the sum of two parts $E_{x,y}^{tot} = E_{x,y}(x,t) + E_{x,y}^{dr}(x,t)$,
the consistent field, satisfying the Maxwell
equations, and the externally applied driving field, respectively.
Moreover, the magnetic field is also the sum of the constant
background field and of the self-consistent field, 
$ B_z^{tot} = B_0 + B_z(x,t) + B_z^{dr}(x,t)$.
The self-consistent fields satisfy the
Maxwell equations:
\begin{equation}
{{\partial E_x}\over{\partial t}} = - j_x , \;\; 
{{\partial E_y}\over{\partial t}} = 
- {{\partial B_z}\over{\partial x}} - j_y ,
\label{3}
\end{equation}
where the two relevant components of the current density are
$j_x = Z n_i V_{ix} - n_e V_{ex}$
and
$j_y = Z n_i V_{iy} - n_e V_{ey}$.
Finally, Poisson equation (used as a check in the code) takes the form
\begin{equation}
{{\partial E_x}\over{\partial x}} = \rho ,
\label{4}
\end{equation}
where the charge density is $\rho = Z n_i - n_e.$
The driving electric field is modelled as 
\[
E_x^{dr}(x,t) = \epsilon_1 {\cal A} ; \;\;\; 
E_y^{dr}(x,t) = \epsilon_2 {\cal A} ; \;\;\; 
B_z^{dr}(x,t) = \epsilon_2 {\cal A} ;  \;\;\;  \;\;\; \;\;\; 
{\cal A} = \delta(x - x_0) \, e^{- t^2/\tau^2} \, \sin{\omega t}
\]
 
where $\epsilon_{1,2}$ are constant amplitudes, $x_0 \in [0,L]$ 
is the position where the driving field acts (usually, $x_0 = 0$), 
$\omega$ is the pump frequency. The
disturbance is switched on and off continuously, with a typical time
scale $\tau = 44.7$. This model allows one to perturb the system at
one boundary of the range $[0,L]$  either 
by an electromagnetic ($\epsilon_1 = 0, \epsilon_2 \ne 0$) or 
by an electrostatic ($\epsilon_1 \ne 0, \epsilon_2 = 0$)
disturbance. The consistent polarization and wave vector are then
defined by the kinetic plasma response, which comprises the nonlinear
coupling
between particle motion and fields contained in the Vlasov Eq.(\ref{1}).

\section{The results of numerical experiments}

\begin{figure}[!t]
\hspace*{-0.2cm}\centerline{\psfig{figure=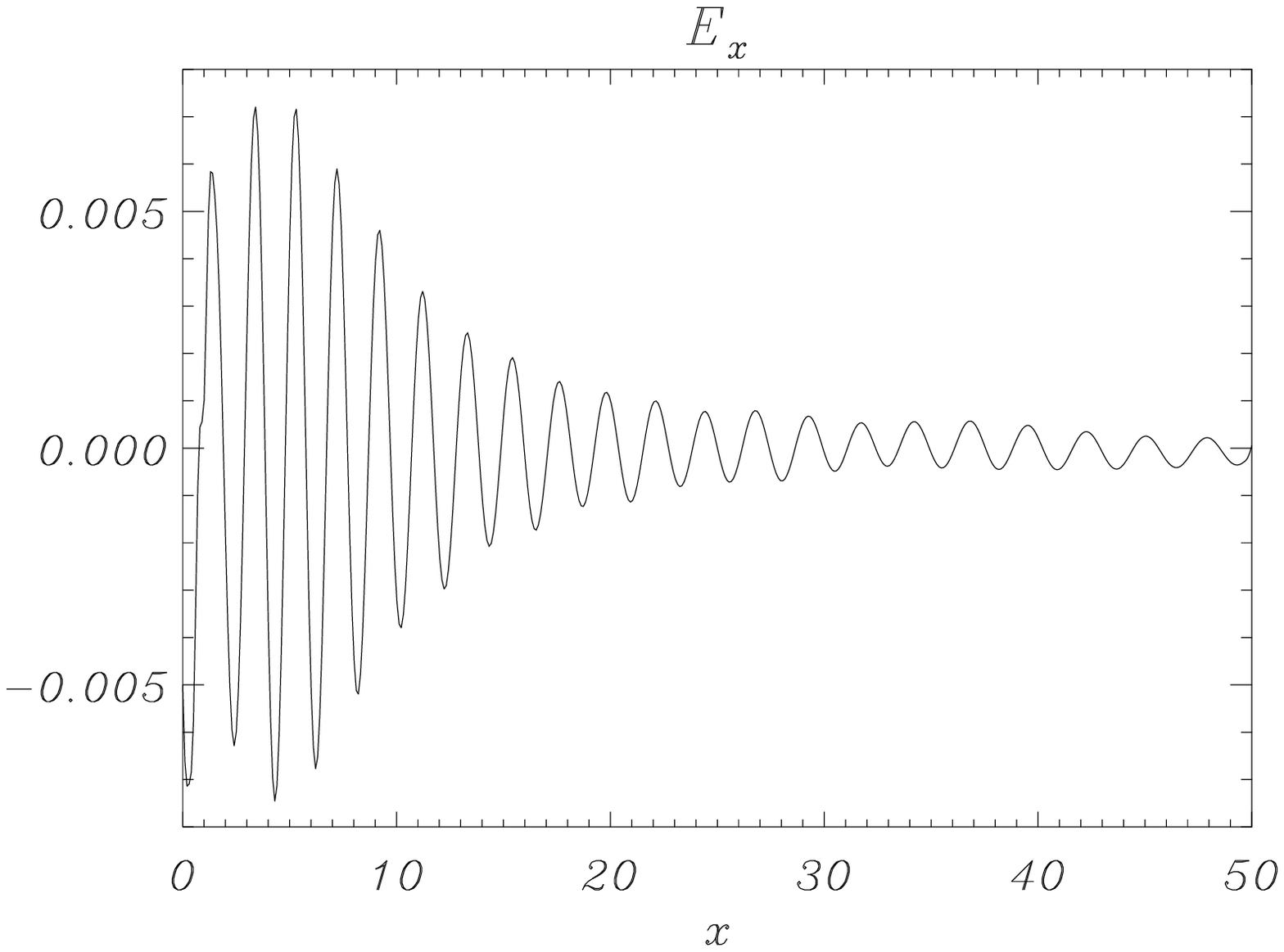,height=8cm,width=8.cm}
\hspace*{-0.5cm}\psfig{figure=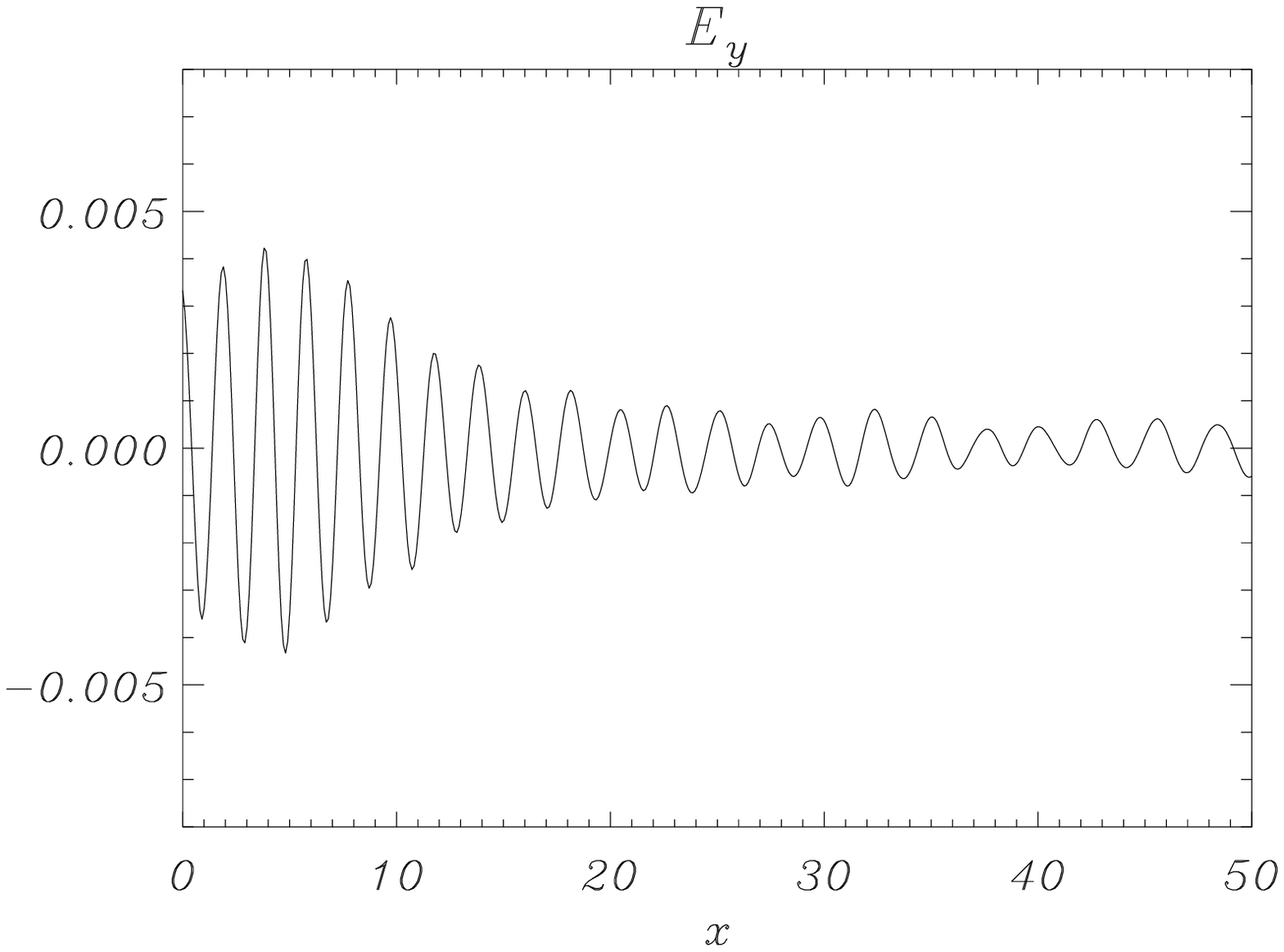,height=8cm,width=8.cm}}
\caption[ ]{The longitudinal, $E_x$, and transversal, $E_y$ components of 
the electric field for $\omega = 2.1$.}
\label{fig2}
\end{figure}

\begin{figure}[!h]
\centerline{\psfig{figure=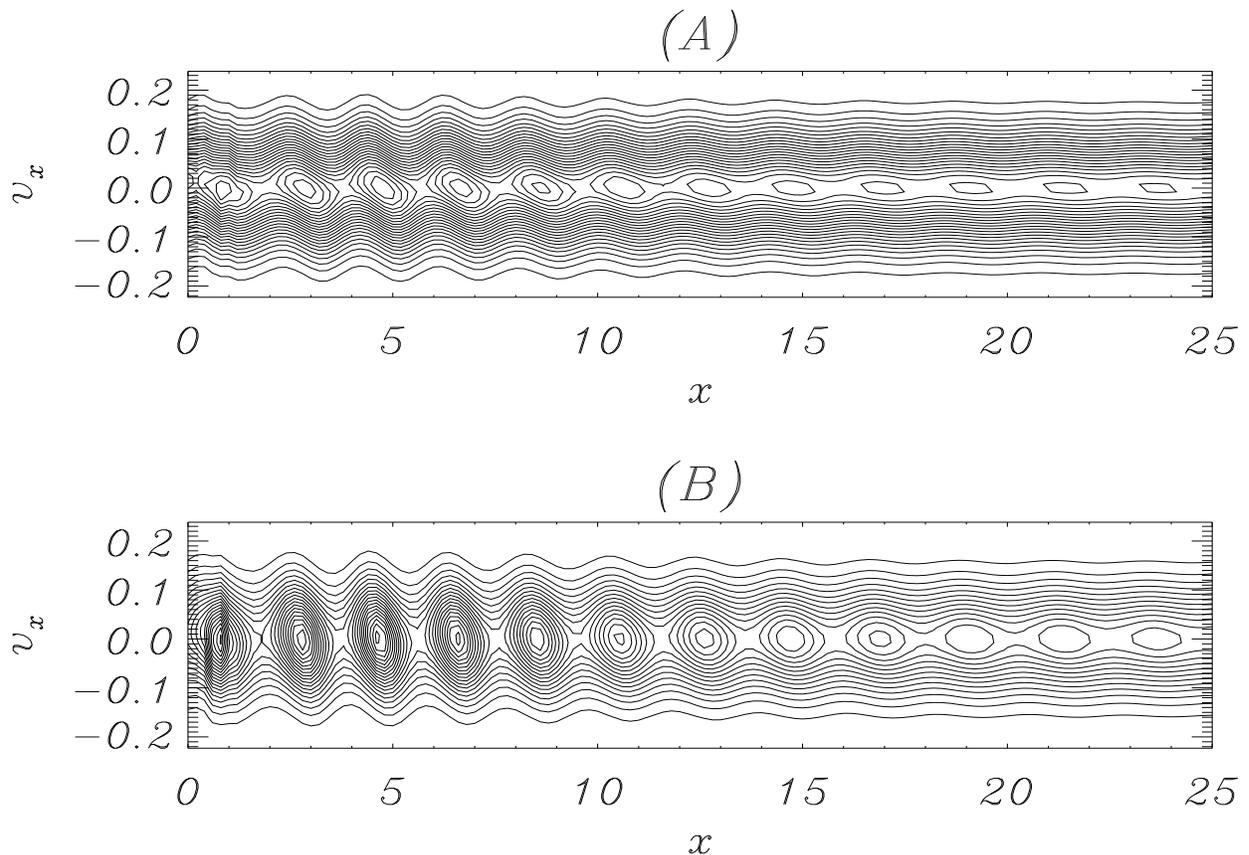,height=12.cm,width=18.cm}}
\caption[ ]{The electron distribution function at $t=100$ in 
the ($x, v_x$) phase space with $v_y = -0.038$, frame (A), and 
$v_y = -0.19$, frame (B).}
\label{fig3}
\end{figure}

Tests aimed at reproducing the wave-plasma interaction in the
low-amplitude (linear) regime have been performed by injecting a 
"pure" electromagnetic wave (i.e. $E_x^{dr} = 0$) at the left 
boundary, $x = 0$, with $\epsilon_2= 0.005$, or by exciting 
an electrostatic perturbation (i.e. $E_y^{dr} = B_z^{dr} = 0$) 
nearby the left boundary, $x \simeq 0$, using a normalized amplitude 
$\epsilon_1 =0.01$. In all cases $v_{th,e}/c = 0.14$, 
which corresponds in dimensional units to
$|E_{x,y}|^2/(4 \pi n_0 T_e) = 5 \times 10^{-3}$.
In dimensionless units the magnetic field is equal to the electron
cyclotron frequency and the value $B = \Omega_{ce} = 2$ has been
chosen in all simulations.
Two values of the pump frequency have been considered: 
$\omega = 0.95$ {\it (a)} and $\omega = 2.1$ {\it (b)}. 
Note that the upper hybrid frequency 
$\omega_{uh} = \sqrt{1 + B^2} = 2.24$ and the upper
cutoff $\omega_{co} = B/2 + \sqrt{1 + B^2/4} = 2.41$.
According to the linear theory of cyclotron waves
\cite{brambilla98}
the cold branches of the dispersion relation of an extraordinary mode
propagating perpendicularly to the magnetic field have a
dimensionless wave vector
$k \approx 3.2$ and $k \approx 0.94$ for two chosen
frequencies, 2.1 and 0.95, respectively. As it is seen in 
Fig. \ref{fig1} the {\it k}-spectra of the electric field manifest a sharp
maximum around the corresponding wave vectors, together with other
features at higher {\it k}'s, independently of the method of
excitation (electrostatic or electromagnetic).
Moreover, since we are dealing with a hot plasma, we should expect
also a component at higher {\it k}, due to the conversion of the
incoming electromagnetic energy into electrostatic 
electron Bernstein waves. In this respect
the most interesting spectrum is that of case {\it (b)}, since the
$\Omega_{ce} < \omega < \omega_{uh}$, where modes with 
$k \approx 10.8$ should be excited. However, in this preliminary 
work where we deal with "small" amplitude external pumps and relatively 
short time of propagation, this effect is not seen. 
In Fig. \ref{fig2} we plot the longitudinal and transversal  
components of the electric field, first and second frame, 
respectively, in the case of an electromagnetic external driver 
of amplitude $\epsilon_2 = 0.005$ and frequency $\omega = 2.1$. 
We see that near the left boundary the energy is transferred to 
the electrostatic counterpart and then, after propagating to 
the right a few $d_e$ at $x \simeq 20$, the amplitudes 
become nearly constant with $E_x \simeq E_y$, despite 
in we are in the limit $k^2 c^2 \gg \omega_{pe}$ where 
the electromagnetic (applied) field should dominate. 
We think that this is due to the warm plasma response. 
In Fig. \ref{fig3} we draw the isolines of the electron distribution 
function at $t=100$ in the ($x, v_x$) phase space at fixed $v_y$ velocity, 
namely $v_y = -0.038$, first frame, and $v_y = -0.19$, second frame (B). 
This figure shows the rotation of the particles around the magnetic 
field corresponding vortex like structures of typical size of 
the order of the selected $v_y$ value. 

In the present analysis the value $v_{th,e}/c = 0.14$ has been chosen, 
which corresponds to an electron temperature of $T_e \approx 10 keV$. 
It is well known that in order to describe correctly the collisionless 
cyclotron waves propagating perpendicularly to the magnetic field, 
the velocity dependence of the electron mass should be considered. 
The propagation of cyclotron waves is however well 
described already by the non relativistic theory. Moreover,
for the frequency values which have been considered,
that is $\omega = 0.95$ and $2.1$ even in the relativistic case,
no appreciable collisionless damping is expected.

\section{Conclusions}
In this work the Vlasov equation for electrons,
coupled with the Maxwell equations, has been numerically solved in an open plasma slab,
where the source of the electromagnetic fields is localized at one of the two spatial boundaries.
The magnetized plasma responses to both purely electromagnetic or electrostatic excitations,
at small amplitudes, have been
studied. The field disturbance is applied in the form of a purely transverse propagating EM wave
or, alternatively, as a purely ES field.
Then, the injected EM energy goes into the longitudinal and transverse components
of the field, almost independently on the excitation details,
and the wavevectors are generated consistently during the wave propagation.
It is seen that, even in the case of a purely EM excitation,
a large fraction of the injected energy can go into the ES field.

The code is particularly suitable for exploring 
the non linear stage of the wave-plasma interaction.
The application of the code to large amplitude driving fields is under way.

\vspace*{-.5cm}

\acknowledgments
One of us, FC, is glad to acknowledge the Plasma Physics Institute 
(IFP) of Milan for supporting in part the research activity on 
the kinetic study of electrostatic and electromagnetic waves 
propagation in a plasma.

\end{document}